%% file: v2.tex
\documentclass[a4paper,german,12pt]{elsart}

\usepackage{amsmath}
\usepackage{amssymb}
\usepackage{wasysym}
\usepackage{bibentry}
\usepackage{epsfig}
\usepackage{epsf}
\usepackage[dvips]{color}
\usepackage{graphicx,graphics}
\usepackage{float}
\usepackage{natbib}
\usepackage{setspace,lineno}
\begin{document}
\linenumbers

\begin{frontmatter}

\title{Warming the early Earth - CO$_2$ reconsidered}

\author[DLR]{\corauthref{cor}Philip von Paris},
\corauth[cor]{Corresponding author: philip.vonparis@dlr.de}
\author[DLR,TU]{Heike Rauer},
\author[DLR,TU]{J. Lee Grenfell},
\author[TU]{Beate Patzer},
\author[DLR]{Pascal Hedelt},
\author[DLR]{Barbara Stracke},
\author[IWF]{Thomas Trautmann} and
\author[IWF]{Franz Schreier}

\address[DLR]{Institut f\"{u}r Planetenforschung, Deutsches Zentrum f\"{u}r Luft- und Raumfahrt, Rutherfordstr.
2, 12489 Berlin (Germany)}

\address[TU]{Zentrum f\"{u}r Astronomie \& Astrophysik, Technische Universit\"{a}t Berlin, Hardenbergstr.
36, 10623 Berlin, (Germany)}

\address[IWF]{Institut f\"{u}r Methodik der Fernerkundung, Deutsches Zentrum f\"{u}r Luft- und Raumfahrt, M\"{u}nchener Str. 20, 82234 Wessling (Germany)}

\begin{abstract}

Despite a fainter Sun, the surface of the early Earth was mostly
ice-free. Proposed solutions to this so-called "faint young Sun
problem" have usually involved higher amounts of greenhouse gases
than present in the modern-day atmosphere. However, geological
evidence seemed to indicate that the atmospheric CO$_2$
concentrations during the Archaean and Proterozoic were far too low
to keep the surface from freezing. With a radiative-convective model
including new, updated thermal absorption coefficients, we found
that the amount of CO$_2$ necessary to obtain 273 K at the surface
is reduced up to an order of magnitude  compared to previous
studies. For the late Archaean and early Proterozoic period of the
Earth, we calculate that CO$_2$ partial pressures of only about 2.9
mb are required to keep its surface from freezing which is
compatible with the amount inferred from sediment studies. This
conclusion was not significantly changed when we varied model
parameters such as relative humidity or surface albedo, obtaining
CO$_2$ partial pressures for the late Archaean between 1.5 and 5.5
mb. Thus, the contradiction between sediment data and model results
disappears 
for the late Archaean and early Proterozoic.

\end{abstract}

\begin{keyword}

Faint Young Sun problem, Earth - Atmospheres, composition -
Radiative transfer

\end{keyword}

\end{frontmatter}

\section{Introduction}

Geological evidence has shown that liquid water was present on the
Earth's surface earlier than 3.7 Gy ago (e.g.,
\citealp{moj1996,rosing2004}) which implies average temperatures on
the surface above 273 K. Some authors have even argued for a hot
Archaean climate ($T$ $>$340 K), based on oxygen \citep{knauth2003}
and silicon \citep{robert2006} isotope analysis of seawater cherts.
However, as pointed out by, e.g., \citet{kasthoward2006} and
\citet{shields2007}, these isotopic signatures changes might not
only be caused by temperature effects. \citet{sleep2006}, for
example, deduced a more moderate surface temperature below 300 K,
based on quartz weathering records in paleosols. Nevertheless, it is
generally accepted that the Earth has been ice-free throughout most
of its history.

Observations of several solar-type stars of different ages and
virtually all standard models of the solar interior have shown that
the total solar luminosity has increased since the ZAMS (Zero Age
Main Sequence) by about 30\% \citep{gough1981,caldeira1992}. Had the
composition of the Earth's atmosphere been the same then as today,
the reduced solar flux would have resulted in surface temperatures
below 273 K prior to 2.0 Gy \citep{sagan1972}. This apparent
contradiction between solar evolution models, climatic simulations
and geological evidence for liquid water and moderately warm
temperatures on Earth has been termed the "faint young Sun problem".


Numerous studies have attempted to solve this problem. For example,
\citet{minton2007} explored the hot early Sun scenario for a
non-standard solar evolution. \citet{shaviv2003} showed that
moderate greenhouse warming in combination with the influence of
solar wind and cosmic rays on climate could resolve the problem.
\citet{jenkins2000} assumed high obliquities in a General
Circulation Model to account for high Archaean temperatures.

However, the most accepted scenario involves a much enhanced
greenhouse effect (GHE) on the early Earth compared to modern Earth.
Today, the GHE produces around 30 K of warming, raising the mean
surface temperature of the Earth to about 288 K. Increased
abundances of greenhouse gases such as carbon dioxide
\citep{kasting1987}, methane \citep{pavlov2000}, ethane
\citep{haqq2008} or ammonia \citep{sagan1972,sagan1997} will
strengthen the GHE, hence potentially resolve the problem. However,
all of these studies faced some form of contradictions or large
uncertainties, either from geological data on atmospheric conditions
or from atmospheric modeling. The formation and destruction of
ammonia is highly dependent on UV levels in the atmosphere
\citep{sagan1997,pavlov2001}. The hydrocarbon haze necessary to
allow higher hydrocarbons to accumulate in the atmosphere depends
critically on the CO$_2$/CH$_4$ ratio \citep{pavlov2003}. The high
values of methane required to heat the surface of the early Earth
depend on estimates of the early biosphere and volcanic activity,
which is not well determined. Past CO$_2$ concentrations required by
atmospheric models \citep{kasting1987} to reach surface temperatures
above 273 K are in conflict with inferred concentrations from the
sediment data \citep{hessler2004,rye1995}.

In this work, the role of CO$_2$ in warming the early Earth is
reconsidered. We used a one-dimensional radiative-convective model,
including updated absorption coefficients in its radiation scheme
for thermal radiation. The model was applied to the atmosphere of
the early Earth in order to investigate the effect of enhanced
carbon dioxide on its climate. Additionally, we investigated the
effect of two important parameters, namely the surface albedo and
the relative humidity, upon the resulting surface temperature.

Our results imply that the amount of CO$_2$ needed to warm the
surface of early Earth might have been over-estimated by previous
studies. Furthermore, the results show that the contradiction
between modelled CO$_2$ concentrations and measured values might
disappear by the end of the Archaean.

Section \ref{describemodel} describes the model and section
\ref{validmracsubsection} the runs to validate the new radiation
scheme. The runs performed for this work are explained and
summarized in section \ref{about}. In section \ref{showresults}, the
results are presented and discussed. Section \ref{conclusions} gives
the summary of the results.

\section{Atmospheric Model}

\label{describemodel}

We used a one-dimensional radiative-convective model based on the
climate part of the model used by \citet{Seg2003} and
\citet{Grenf2007a,Grenf2007b}. Our model differs in upgrades of the
radiation scheme to calculate the thermal emission in the
atmosphere. The model calculates globally, diurnally-averaged
atmospheric temperature and water profiles for cloud-free
conditions. We will first state some basic characteristics of the
model (\ref{basic}). Then we will describe the calculation of the
energy transport via radiative transfer (solar and thermal fluxes)
and convection (\ref{tempprofcalc}) to obtain the atmospheric
temperature profile. Thereafter, a description of the determination
of the water profile is given (\ref{watcalcu}). Finally, the model
input parameters are summarized (\ref{summarymodelinput}).

\subsection{Basic model description}

\label{basic}

The model determines the temperature profile by assuming two
dominant mechanisms of energy transport, i.e. radiative transfer and
convection. The convective lapse rate is assumed to be adiabatic.
The radiative lapse rate is calculated from contributions of both
solar and thermal radiation, including Rayleigh scattering for solar
radiation and continuum absorption in the thermal region. Table
\ref{thermalinput} summarizes the contributions of the different
atmospheric species to the calculation of the temperature profile.


The species considered in the model are molecular nitrogen, water,
molecular oxygen, argon, carbon dioxide and carbon monoxide. For
example, a typical early Earth run considered 0.77 bar of nitrogen
and several (2.9 - 57.2) mb of carbon dioxide in addition to water
in varying concentrations (0.5---1 \% at the surface). Molecular
nitrogen is an effective Rayleigh scatterer (although not as
effective as carbon dioxide) and as a main constituent of the
atmosphere also contributes to the heat capacity. Water is not
considered to be an important Rayleigh scatterer, but it is relevant
for the other radiative processes. Also, water influences the
adiabatic lapse rate because it readily condenses in the
troposphere. However, due to small mixing ratios, especially in the
stratosphere, water vapour does  not contribute to the heat capacity
of the atmosphere. Molecular oxygen and argon do contribute to the
heat capacity of the atmosphere, and molecular oxygen additionally
contributes to the Rayleigh scattering coefficient. Carbon dioxide
contributes to all relevant radiative mechanisms (molecular
absorption of solar and thermal radiation, continuum absorption,
Rayleigh scattering), but not to the adiabatic lapse rate because it
does not condense under conditions described in this paper. Carbon
monoxide is an important absorber species in some mid-infrared
windows and contributes to the heat capacity.

The model assumes the hydrostatic relation between pressure $p$ and
density $\varrho$ throughout the plane-parallel atmosphere. On the
52 model layers, a logarithmic pressure grid is calculated.
Specified pressure levels at the planetary surface (e.g., 1 bar for
the standard Earth case) and the upper model lid (at
6.6$\cdot10^{-5}$ bar) determine the altitude range which, for
modern Earth conditions, extends to 65-70 km, i.e. the lower- to mid
mesosphere. For all gases except water, the ideal gas law is taken
as the equation of state (see \ref{convadjust} for water). The
effect of clouds is difficult to incorporate into 1D models (see,
for example, \citealp{pavlov2000,Seg2003}). In the present model,
following the approach of \citet{Seg2003}, clouds are implicitly
included by adjusting the surface albedo $A_{\rm{surf}}$ such that
under modern Earth control conditions the model calculates a mean
surface temperature of 288 K. The required value for $A_{\rm{surf}}$
is about 0.21, whereas the actual global value for Earth is
approximately 0.15. This can be interpreted as a ground cloud layer
instead of tropospheric or stratospheric clouds. The model uses a
time-stepping algorithm to convergence to the steady-state
solution\citep{pavlov2000}.

\subsection{Temperature profile}
\label{tempprofcalc}

During each time step, the temperature profile is calculated from
the radiative equilibrium condition. The temperature $T$ on an
atmospheric level $z$ is determined by the following equation of
energy conservation \citep{pavlov2000}:

\begin{equation}\label{timestep}
    \frac{d}{dt}T(z)=-\frac{g}{c_p(T,z)}\frac{dF(z)}{dp(z)}
\end{equation}

where $dt$ is the time step in the model, $c_p$ the heat capacity,
$F$ the total net radiative flux and $p$ the pressure of the level.
The radiative flux $F$ is the sum of thermal planetary and
atmospheric emission, $F_{\rm{thermal}}$, and the solar radiative
input, $F_{\rm{solar}}$, into the atmosphere:

\begin{equation}
F(z)=F_{\rm{thermal}}(z)+F_{\rm{solar}}(z) \label{fluesse}
\end{equation}

These fluxes are calculated separately by two numerical schemes
which solve the monochromatic radiative transfer equation (RTE) for
the spectral intensity $I_{\nu}$ in the respective spectral domain
(i.e., near-UV to near-IR for solar flux, near-IR to far-IR for
thermal flux):

\begin{equation}
\mu\frac{dI_{\nu}}{d\tau_{\nu}}= I_{\nu}-S_{\nu}\label{endgueltig}
\end{equation}

where $S_{\nu}$ is the source function (either the incident solar
radiation or the thermal blackbody emission of the atmospheric
layers and the planetary surface), $d\tau_{\nu}$ the optical depth
and $\mu=\cos(\theta)$ the cosine of the polar angle. The optical
depth is defined as usual by

\begin{equation}\label{opttiefe}
d\tau=-(\kappa_{\nu}+s_{\nu})dz
\end{equation}

where $\kappa_{\nu}$ and $s_{\nu}$ represent the absorption
coefficient and the scattering coefficient respectively. The
absorption coefficient for a gas mixture is calculated from the
individual absorption coefficients of the gas species $i$:

\begin{equation}\label{kappadefinition}
   \kappa_{\nu}=\sum_i \kappa_{\nu,i}=\sum_i \sigma_{\rm{abs},i} \cdot N_i
\end{equation}

where $N_i$ is the number density and $\sigma_{abs,i}$ the molecular
absorption cross section of the gas species $i$. When no scattering
occurs (i.e., $s_{\nu}=0$), eq. \ref{opttiefe} can be written in
terms of the column density $W_i$ of the gas species $i$ as:

\begin{equation}\label{optcolumn}
\tau=\sum_i \sigma_{\rm{abs},i}\cdot W_i
\end{equation}

The absorption cross section is defined by:

\begin{equation}\label{defabs}
    \sigma_{\rm{abs}}(\nu,p,T)=\sum_j S_{j}(T)\cdot g_{j}(\nu,T,p)
\end{equation}

Here, $S_{j}(T)$ is the temperature-dependent line strength of a
particular spectral line $j$ and $g_{j}(\nu,T,p)$ the temperature-
and pressure-dependent line shape function of the same line.

For the scattering coefficient, an analogous equation is valid:

\begin{equation}\label{kappascattdefinition}
    s_{\nu}=\sum_i s_{\nu,i}=\sum_i \sigma_{y,i}(\nu) \cdot N_i
\end{equation}

Here, $\sigma_{y,i}(\nu)$ is a scattering cross section of type $y$.
In the present model, Rayleigh scattering is considered.

The solution of eq. (\ref{opttiefe}), i.e. the calculation of
optical depths, is one of the key elements in radiative-convective
models. This solution, of course, depends on the accurate
calculation of the absorption cross sections.

From eq. (\ref{endgueltig}), the necessary fluxes for eq.
(\ref{fluesse}) (i.e., the thermal and the solar flux) are obtained
by an angular integration of the (monochromatic) intensity:

\begin{equation}
F_{\nu}=\int_{\Omega}^{}d\omega \mu I_{\nu} \label{angular}
\end{equation}

and a frequency integration of the (monochromatic) flux:

\begin{equation}
F_{\nu_1\rightarrow\nu_2}=\int_{\nu_1}^{{\nu_2}}F_{\nu}d\nu
\label{frequ}
\end{equation}

Each one of these integrations is performed independently for the
two components of the total flux.

\subsubsection{Solar radiation}

The solar radiation module which calculates $F_{\rm{solar}}(z)$ for
eq. (\ref{fluesse}) has already been used by, e.g.,
\citet{pavlov2000,Seg2003} or \citet{Grenf2007a} and is based on
\citet{kasting1984} and \citet{kasting1988}. The module considers a
spectral range from 0.26 to 4.5 $\mu$m in 38 intervals. It evaluates
the solar incident radiation at a fixed daytime average solar zenith
angle of 60$^\circ$. Contributions to the optical depth come from
gaseous absorption by water and carbon dioxide (i.e., $\kappa_{\nu}$
in eq. (\ref{opttiefe})) and from Rayleigh scattering by carbon
dioxide, molecular nitrogen and molecular oxygen (i.e., $s_{\nu}$ in
eq. (\ref{opttiefe})).

Absorption cross sections $\sigma_{\rm{abs}}$ for the solar code
were obtained from the HITRAN 1992 database \citep{rothman1992}.
Rayleigh scattering cross sections $\sigma _{ray}$ are parameterized
following \citet{vardavas1984}. The frequency integration (see also
eq. (\ref{frequ})) of the RTE for $F_{solar}$ in each of the 38
spectral intervals is parameterized by a four-term correlated-k
exponential sum (e.g., \citealp{wiscombe1977}). The following
angular integration (eq. (\ref{angular})) is performed by using a
quadrature $\delta$-2-stream approximation code based on
\citet{Toon1989}. The resulting fluxes from each spectral interval
are added up to yield the total solar flux $F_{\rm{solar}}(z)$ at an
atmospheric level $z$. This flux is further multiplied by a factor
of 0.5 to account for diurnal variation.

\subsubsection{Thermal molecular absorption}

\label{MRAC}

The thermal (planetary) radiation module for $F_{\rm{thermal}}(z)$
in eq. (\ref{fluesse}) considers a spectral range from 1 to 500
$\mu$m in 25 intervals. Our new thermal module is called MRAC
(Modified RRTM for Application in CO$_2$-dominated Atmospheres) and
is based on the radiation scheme RRTM (Rapid Radiative Transfer
Model). RRTM was developed by \citet{Mlawer1997} and has been used
by numerous other modeling studies (e.g.,
\citealp{Seg2003,Seg2005,Grenf2007a,Grenf2007b}). The need for a new
radiation model comes from the fact that RRTM was specifically
designed for conditions of modern Earth, i.e. it is not adaptable to
studies of atmospheres which greatly differ from modern atmospheric
conditions (in terms of atmospheric composition, temperature
structure, pressure, etc.). MRAC is easily adaptable to varying
conditions, as is described below.

MRAC uses the correlated-$k$ approach (e.g.,
\citealp{goody1989,lacis1991,Cola2003}) for the frequency
integration of the RTE in the thermal range, as does RRTM. The
planetary surface (bottom layer of the model atmosphere) and the
atmospheric layers are taken as blackbody emitters, according to
their respective temperatures. The thermal surface emissivity is set
to unity. The absorber species considered in MRAC are water, carbon
dioxide and carbon monoxide. The angular integration (eq.
(\ref{angular})) is performed using the diffusivity approximation
(as in \citealp{Mlawer1997}).

\textit{The correlated-k method}

Basically, this method transfers the frequency integration (FI) of
the RTE from frequency space $\nu$ to a probability space $g$. For
the absorption cross section $\sigma_{\rm{abs}}$ of eq.
(\ref{defabs}) in an interval $[\nu_1,\nu_2]$, a probabilistic
density distribution $f(\sigma_{\rm{abs}})$ (i.e., probability of
occurence for a particular value of $\sigma_{\rm{abs}}$) with the
following normalization condition can be defined:

\begin{equation}
\int_{0}^{{\infty}}f(\sigma_{\rm{abs}})d\sigma_{\rm{abs}}=1
\label{fk}
\end{equation}

To follow conventional nomenclature in the literature regarding
$k$-distributions, $\sigma_{\rm{abs}}$ is hereafter referred to as
$k$. The function $f(\sigma_{\rm{abs}})=f(k)$ is called the
$k$-distribution. From the $k$-distribution, a cumulative
$k$-distribution $g(k)$ can be defined by

\begin{equation}
g(k)=\int_{0}^{{k}}f(k')dk' \label{gk}
\end{equation}

The cumulative $k$-distribution is a strictly monotonic function and
may thus be inverted from $g(k)$ to yield $k(g)$. This mapping of
the frequency information ($k(\nu)$) into a single probability
variable ($k(g)$) can be done because it is irrelevant at which
position of the spectral interval a particular value of the
absorption cross section $k$ occurs. Performing a variable
substitution in eq. (\ref{frequ}) then leads to the following
equation:

\begin{equation}
F_{\nu_1\rightarrow\nu_2}=\int_{\nu_1}^{{\nu_2}}F(\nu)d\nu=\int_{0}^{{1}}F(g)dg
\label{corrk}
\end{equation}

The goal of the cumulative-$k$ approach is to reduce the number of
radiative transfer operations drastically while keeping the accuracy
of line-by-line models. This can be achieved with very few numbers
of points in $g$ space (see \citealp{goody1989,west1990}). The $g$
integration in eq. (\ref{corrk}) is performed in MRAC by Gaussian
quadrature using 16 intervals in $g$ space (the same as used in
RRTM, \citealp{Mlawer1997}).

The extension of this exact method from homogeneous to inhomogeneous
atmospheres is called the correlated-$k$ method (e.g.,
\citealp{Mlawer1997}). Each $g$ interval is treated as if it were a
monochromatic frequency interval, i.e. the method uses the same
subset of $g$ space for all layers throughout the atmosphere. This
implies a full frequency correlation of a specific subset of $g$
space for all atmospheric layers. There are conditions under which
this approach is exact \citep{goodyyung1989}, but usually these do
not hold. However, the numerical error of the correlated-$k$ method
is generally small \citep{Mlawer1997}.

\textit{Creating the new radiation scheme}

MRAC was originally designed to simulate atmospheres of a wide range
of possible terrestrial planets other than modern Earth, as stated
above. Therefore, a new temperature-pressure ($T$-$p$) grid to
incorporate the aforementioned ($T$,$p$)-dependence of the
absorption cross sections (see eq. \ref{defabs}) has been
introduced. In RRTM, $k$ values are tabulated for every spectral
band and every point in $g$ space for 59 pressure levels and the
associated Mid-Latitude-Summer (MLS) standard Earth temperature
values as well as temperature values T$_{\rm{MLS}}\pm$ 15 K and
T$_{\rm{MLS}}\pm$ 30 K, as described in \citet{Mlawer1997}. The
$T$-$p$ grid of RRTM is thus more or less fixed to modern Earth
conditions. For MRAC, we used 8 pressure levels, ranging
equidistantly in log $p$ from 10$^{-5}$ to 100 bar and 9 temperature
points, 6 in 50 K steps from 150 K to 400 K and three additional
points for 500, 600, 700 K respectively. Tests showed that this grid
allows for an interpolation accuracy of usually better than 2-3\%.
Furthermore, the number of $T$-$p$ points is consistent with
previous modeling studies \citep{kasting1993,Cola2003}.
\newline
Figure \ref{range} shows the range of the tabulated $k$-values for
both RRTM and MRAC, i.e. the interpolation regime of the two
radiative schemes. It demonstrates that RRTM can only be applied to
a much narrower range of atmospheric conditions in comparison to
MRAC.


The necessary re-calculation of the $k$-distributions for each of
the gases included in MRAC (water, carbon dioxide, carbon monoxide)
proceeded in three steps:

First, the absorption cross sections for every species were
calculated for each $T$-$p$ grid point in each of the spectral
intervals where the respective species is active. The line shape
cut-off was set to 10 cm$^{-1}$ from the frequency $\nu$, i.e. the
sum over $j$ in eq. (\ref{defabs}) contains contributions from all
lines within $\pm$ 10 cm$^{-1}$. For the line shape
$g_{j}(\nu,T,p)$, a Voigt profile was assumed. The required line
parameters were taken from the HiTemp 1995 database
\citep{rothman1995}. The foreign broadening parameters in HiTemp are
given for air, i.e. an oxygen-nitrogen mixture as a background
atmosphere. However, as reported by several authors (e.g.,
\citealp{brown2005,toth2000}), the foreign broadening parameters
vary by significant amounts when different broadening gases are
considered.  Thus, for each type of background atmosphere, a new set
of $k$-distributions must be generated (e.g., low-oxygen
atmospheres, CO$_2$-dominated atmospheres, intermediate
N$_2$-O$_2$-atmospheres). These line parameters were then used as
input to a line-by-line radiative transfer model called Mirart
\citep{schreier2003}. Mirart produced the actual absorption cross
sections with a spectral resolution of 10$^6$ equidistant points per
spectral interval.

Second, the $k$-distributions $f(k)$ were calculated from the
absorption cross sections. From the $k$ distribution $f(k)$, the
cumulative $k$-distribution $g(k)$ was then obtained.

Third, representative $k$ values were calculated for each of the $g$
subintervals. In this step, our algorithm followed the approach of
\citet{Mlawer1997}, i.e. for each of the 16 Gaussian subintervals in
$g$ space, an arithmetic mean absorption cross section was
calculated.

MRAC also implements a so-called binary species parameter $\eta$ for
transmittance calculations. This is employed in intervals having two
important absorber species (for more details, see
\citealp{Mlawer1997} or \citealp{Cola2003}):

\begin{equation}\label{bsp}
\eta=\frac{C_1}{C_1+r\cdot C_2}
\end{equation}

Here, $C_{1,2}$ are the concentrations of the two gases (in the case
of MRAC, water and carbon dioxide) and $r$ is some specified
reference ratio (mean modern Earth tropospheric values). Carbon
monoxide, although present in the model in six spectral intervals,
is not considered to be part of the binary species parameter. This
is partly due to the expected low concentrations in the simulated
atmospheres (typical theoretical and measured values for early Earth
and Mars are below 10$^{-4}$ volume mixing ratio), partly due to the
expected low temperatures (below 300 K), which means that the strong
CO fundamentals around 4 $\mu$m are completely outside the relevant
Planck emission windows. Consequently, carbon monoxide has only a
reduced impact on the radiation budget, compared to water and carbon
dioxide. $k$-distributions are calculated in MRAC for 5 different
values of $\eta$, ranging equidistantly from 0 (CO$_2$ only) to 1
(H$_2$O only).

Another improvement in MRAC, compared to RRTM, is the treatment of
the Planck function in each band. The fraction of thermal radiance
associated with a subset in $g$ space is calculated from eq. (11) in
\citet{Mlawer1997}:

\begin{equation}\label{rrtmfg}
    f_g=\frac{B_gw_g}{\overline{B}_g}
\end{equation}

Here, $B_g$ is the average Planck function of the frequencies in the
subset of $g$ space, $w_g$ the Gaussian weight of the $g$ interval
and $\overline{B}_g$ the average Planck function of the whole
spectral band (i.e., the whole $g$ space). As temperature, pressure
and species concentrations vary, the different $g$ subsets will
correspond to different frequencies, and as such the value of $B_g$,
thus $f_g$, will vary. This has been taken into account while
constructing the $k$-distributions.

\citet{Mlawer1997} tabulated values of $f_g$ for every value of the
binary species parameter and for two atmospheric reference levels,
one each in the troposphere and the stratosphere. In MRAC, values of
$f_g$ were tabulated for three temperatures and two pressure levels
as well as for the values of the binary species parameter. The most
important factor for $f_g$ is the binary species parameter $\eta$,
whereas the variation with temperature and pressure is rather small,
although not negligible. Therefore these 6 $T$-$p$ points are
regarded as to be sufficient.

A further difference between RRTM and MRAC is the distinction
between troposphere and stratosphere. In the troposphere, RRTM
changes major and minor absorbers in some of the spectral intervals
(see table 1 in \citealp{Mlawer1997}). In some spectral bands, no
absorption is considered in the stratosphere, in others, the number
of key species is reduced. This is done because on Earth, the
chemical and physical regimes are quite different in the troposphere
compared to those in the stratosphere. However, as this is mostly
due to Earth-specific conditions (e.g., the cold trap, tropopause
and temperature inversion all approximately occur at the same
altitude), this distinction was not incorporated into MRAC.

\subsubsection{Thermal continuum absorption}


Based on approximation formulations used by \citet{kasting1984},
\citet{kasting1984water} and \citet{Cola2003}, additional CO$_2$ and
H$_2$O continuum absorption in the thermal region is considered. In
contrast, the RRTM scheme only considers water continuum absorption
\citep{Mlawer1997}.

Equation (\ref{co2contkasting}) shows the approximation for the
optical depth $ \tau_{\rm{cont},\rm{CO_2}}$ due to CO$_2$ continuum
absorption. The corresponding parameters are taken from
\citet{kasting1984}.

\begin{equation}\label{co2contkasting}
    \tau_{\rm{cont},\rm{CO_2}}=C_iW\cdot p_E\left(\frac{T_0}{T}\right)^{t_i}
\end{equation}

In this equation, $C_i$ a frequency-dependent adjustment to the path
length, $W$ the column amount of CO$_2$, $p_E=(1+0.3\cdot
C_{\rm{CO_2}})\cdot p$ ($p$ layer pressure, $C_{\rm{CO_2}}$
concentration) an effective CO$_2$ broadening pressure and $T_0$=300
K is a reference temperature.

The optical depth $\tau_{\rm{cont},\rm{H_2O}}$ due to water
continuum absorption in the window region (8-12 $\mu$m) is
calculated from the equation \citep{kasting1984water}

\begin{equation}\label{waterkont}
\tau_{\rm{cont},\rm{H_2O}}=h_n\cdot p \cdot \frac{W_w^2}{W_t}
\end{equation}

where $h_n=h_t\cdot h_{\nu}$ incorporates the frequency and
temperature dependence, $p$ is the pressure, $W_t$ the total and
$W_w$ the water column of the layer. $h_{\nu}$ is evaluated at the
high frequency interval boundary, as in \citet{kasting1984water}. We
use the following approximations for $h_t$ and $h_{\nu}$, based on
\citet{kasting1984water} and \citet{Cola2003}:

\begin{equation}\label{approxtempwater}
    h_t=e^{1800\cdot(\frac{1}{T}-\frac{1}{296})}
\end{equation}
\begin{equation}\label{approxfreqwater}
    h_{\nu}=1.25\cdot 10^{-22}+1.67\cdot 10^{-19}\cdot e^{-2.62\cdot 10^{-13}\cdot \nu}
\end{equation}

Both the water and the carbon dioxide continuum absorption are
considered to be approximately monochromatic over a specific
spectral interval, hence their contribution to the overall
absorption coefficient (see eq. (\ref{opttiefe})) is added as a
constant term.

\subsubsection{Convective adjustment}

\label{convadjust}

Convective adjustment to the lapse rate is performed whenever the
calculated radiative lapse rate $\nabla_{\rm{rad}}T$ exceeds the
adiabatic value $\nabla_{\rm{ad}}T$ (Schwarzschild criterion):

\begin{equation}
\nabla_{\rm{rad}}T>\nabla_{\rm{ad}}T\label{schwarz}
\end{equation}

The adiabatic lapse rate is calculated as a standard dry adiabat in
the stratosphere. In the troposphere, a wet H$_2$O adiabatic lapse
rate is assumed. Below 273 K, the Clausius-Clapeyron-equation
$\frac{d \ln(p_v)}{d \ln(T)}=\frac{m\cdot L}{R\cdot T}$ ($R$
universal gas constant, $m$ mass, $L$ latent heat release per mass)
for the saturation vapor pressure curve $p_v$ is applied. Between
273 and 647 K, a formulation by \citet{ingersoll1969} is taken.

\subsection{Atmospheric water profile}

\label{watcalcu}

In every time step, the water vapor profile is re-calculated
according to the new temperature profile.

In the troposphere, water vapor concentrations $C_{\rm{H_2O}}$ are
calculated from a fixed relative humidity distribution $RH$:

\begin{equation}\label{relhumwater}
C_{\rm{H_2O}}(T,z)=\frac{p_{\rm{sat}}(T(z))}{p(z)}\cdot RH(z)
\end{equation}

where $p_{\rm{sat}}$ is the saturation vapor pressure of water at
the given temperature $T$ and $p$ the atmospheric pressure at level
$z$. The default relative humidity profile $RH$ follows the approach
of \citet{manabewetherald1967}, with a relative humidity $R_s$ of
80\% at the surface.

\begin{equation}\label{manabewetherald}
RH(z)=R_s \cdot \frac{\frac{p(z)}{p_{\rm{surface}}}-0.02}{0.98}
\end{equation}

Above the cold trap, water vapor is treated as a non-condensable
gas, and its concentration is fixed at the cold trap value.

\subsection{Boundary conditions, initial values and parameters}

\label{summarymodelinput}

Since eq. (\ref{timestep}) is a first order differential equation
for the temperature, a starting temperature profile must be
provided. In addition, a boundary condition for the radiative flux
must be specified. To obtain unique equilibrium solutions,
parameters must also be provided for the model. These include
pressure parameters for the planetary top-of-atmosphere (TOA)
pressure $p_0$, gas concentrations, surface albedo or solar zenith
angle, for example.

Table \ref{paraminput} summarizes the boundary conditions, initial
values and parameters.


\section{Validation of the new radiation scheme}

\label{validmracsubsection}

\subsection{General remarks}

MRAC has been tested in two different ways.

\begin{itemize}
  \item Case 1: $k$-distributions
\newline
The calculated $k$-distributions, i.e. the model input data, have
been validated against published values to show that the algorithm
creating the $k$-distributions works correctly.

  \item Case 2: Earth atmosphere temperature profiles
\newline
Temperature profiles of an Earth-like test atmospheres (composition:
N$_2$ 0.77, O$_2$ 0.21, Ar 0.01, CO$_2$ 3.55 $\cdot$ 10$^{-4}$)
calculated with MRAC and RRTM have been compared. This was done
since RRTM has been extensively validated both against line-by-line
codes and atmospheric measurements \citep{Mlawer1997} under modern
Earth conditions.

Our test atmosphere has a composition close to the present day
atmosphere. However, it lacks radiative trace gases such as nitrous
oxide, ozone and methane, as these gases cannot be handled by MRAC
yet (see above). Note that due to the lack of ozone in our test
atmosphere, we do not expect a large stratospheric temperature
maximum as is observed in the present Earth atmosphere because this
maximum is almost entirely due to the absorption of solar radiation
by ozone.

We additionally performed test runs on a second test atmosphere (not
shown) which differs from the first one by its CO$_2$ content
(100-fold increase). This 100-fold increase in CO$_2$ represents the
current limit for the RRTM scheme (\citealp{Seg2003}, Eli Mlawer,
priv. comm.).

\end{itemize}

These two validation approaches are discussed below.

\subsection{$k$-distributions}

\begin{figure}[H]
  \includegraphics[width=400pt]{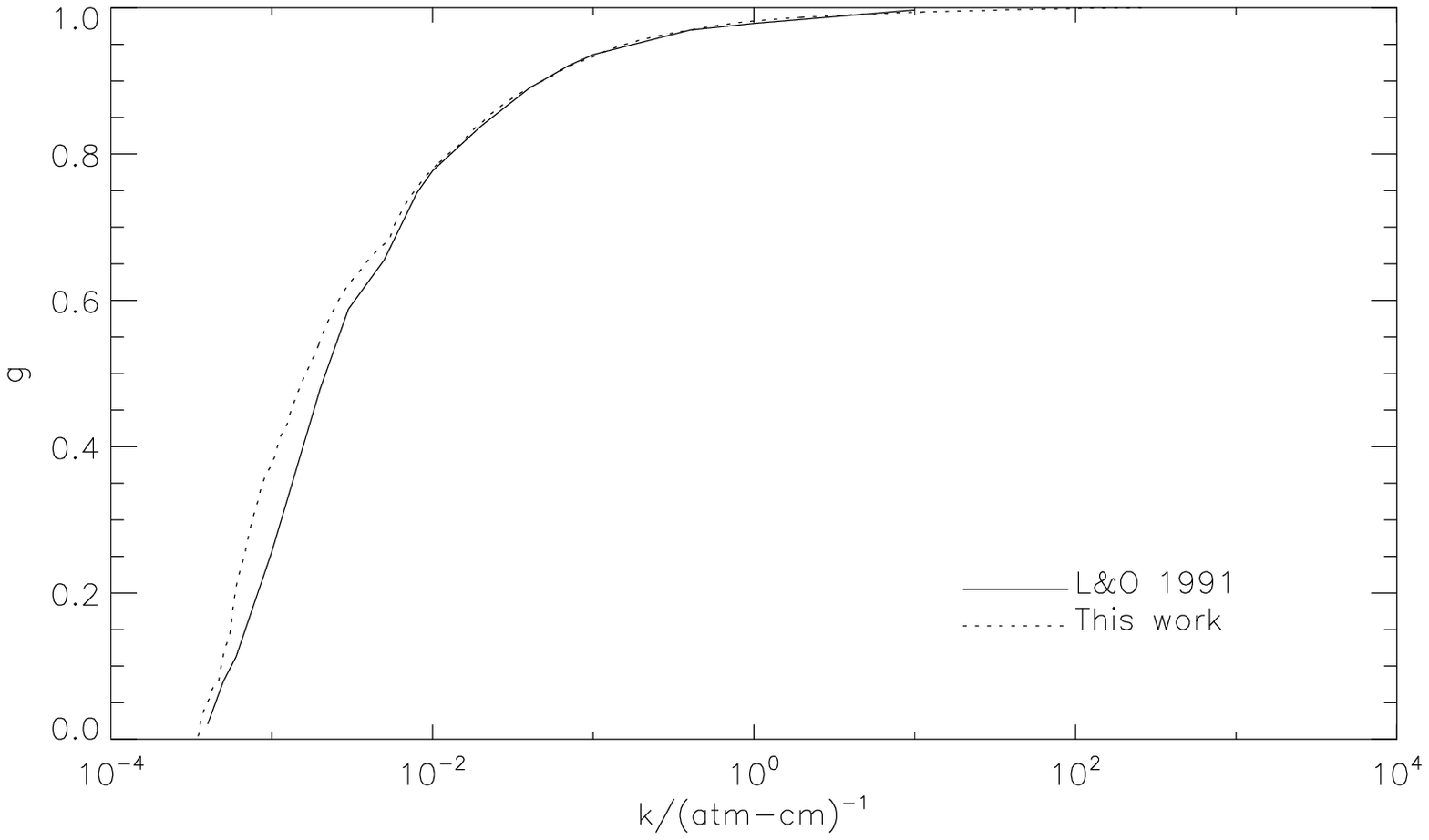}\\
  \caption{}
    \label{kdislacis}
\end{figure}

\begin{figure}[H]
  \includegraphics[width=400pt]{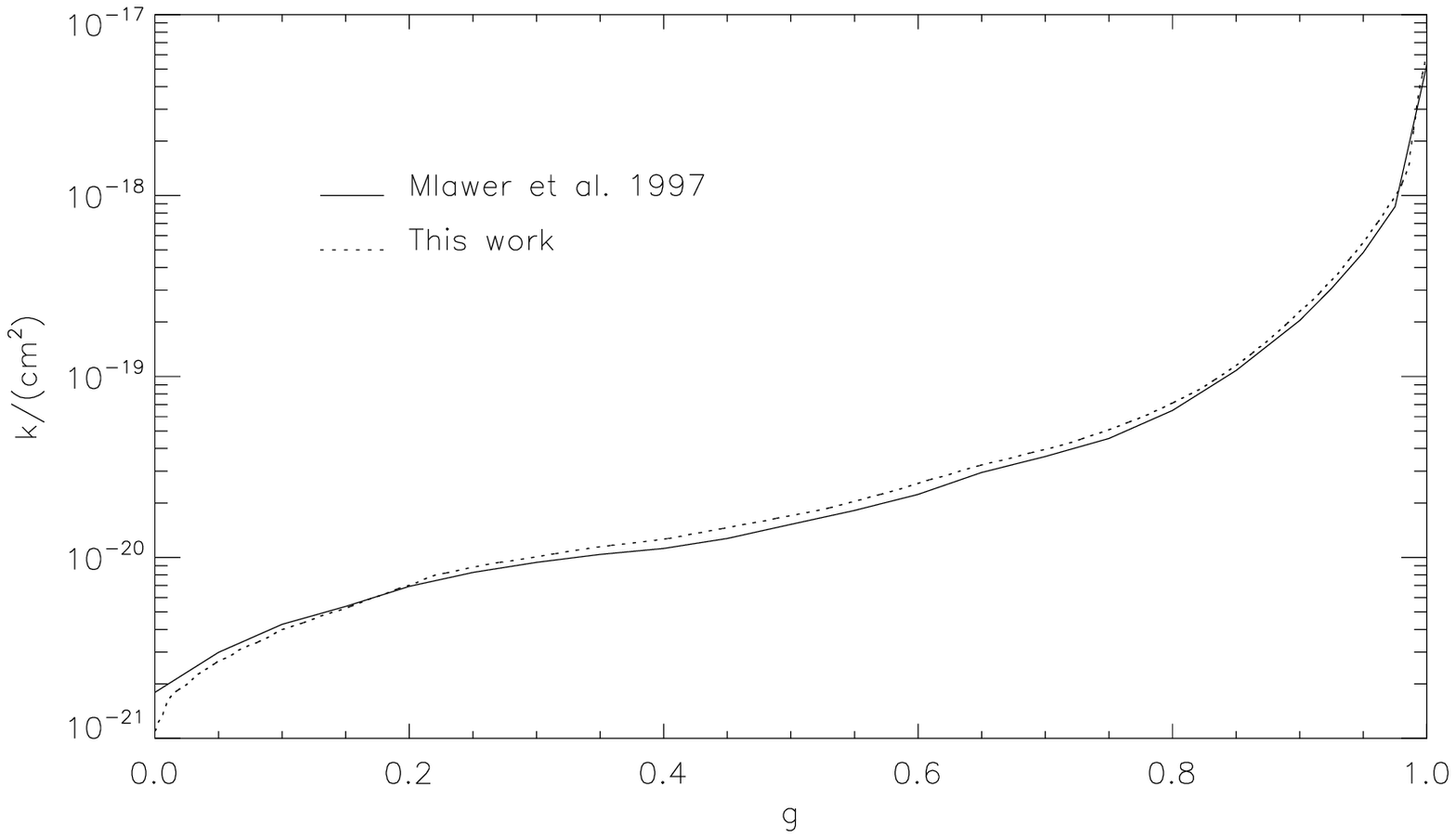}\\
    \caption{}
    \label{kdismlawer}
\end{figure}

Figures \ref{kdislacis} and \ref{kdismlawer} compare our calculated
$k$-distributions (dotted lines) with previously published values
(plain lines) for different water and carbon dioxide bands.
Published values were taken from \citet{lacis1991} (H$_2$O, Fig.
\ref{kdislacis}) and \citet{Mlawer1997} (CO$_2$, Fig.
\ref{kdismlawer}).
Figures \ref{kdislacis} and \ref{kdismlawer} indicate quite good
agreement with the published values.

\subsection{Earth temperature profiles}

\begin{figure}[H]
  \includegraphics[width=470pt]{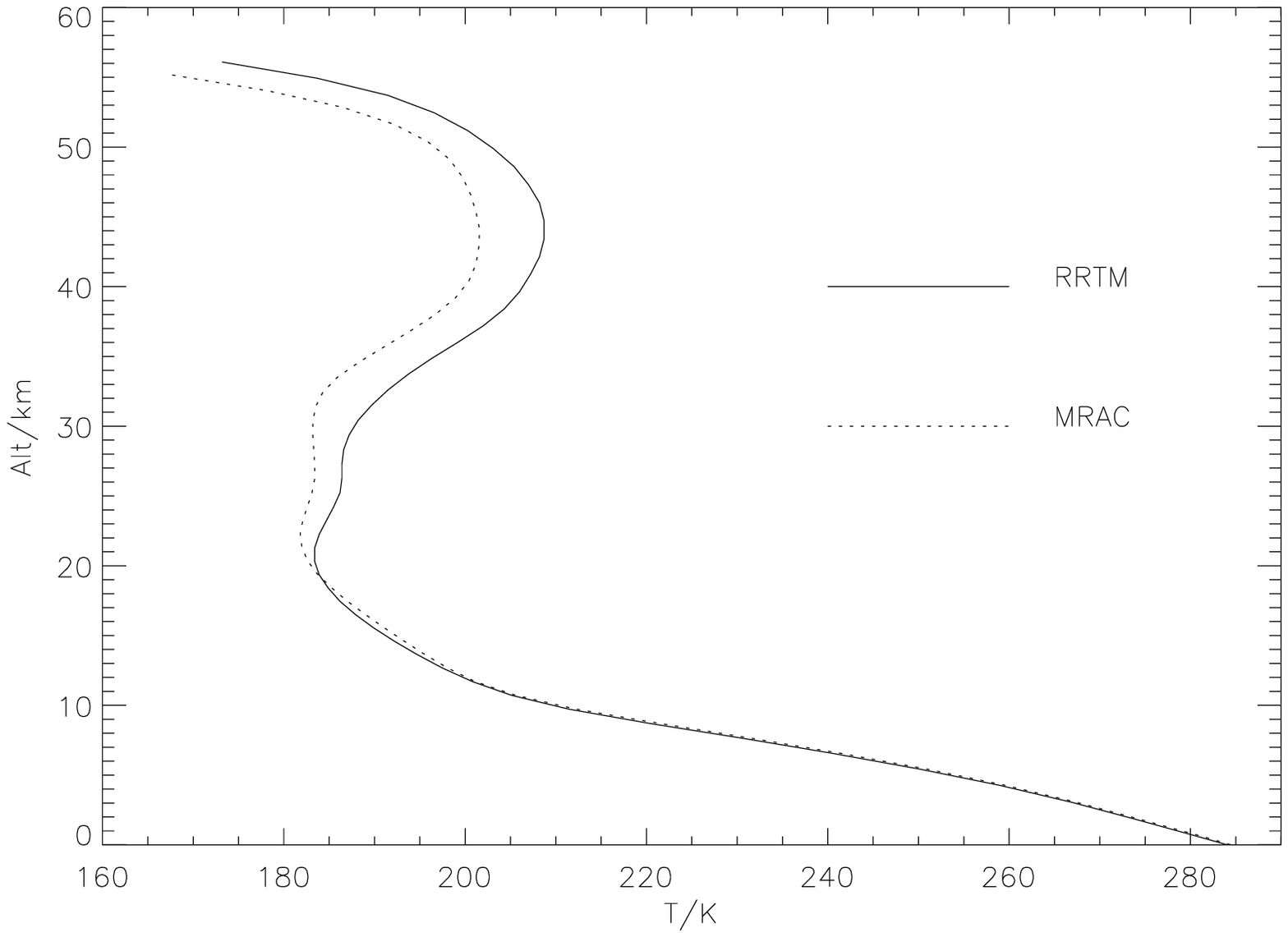}
     \caption{}
  \label{MRACvalid}
\end{figure}

The calculated temperature profiles for the test atmosphere  are
shown in Fig. \ref{MRACvalid}.


Figure \ref{MRACvalid} implies some differences in the middle to
upper stratosphere (2-6 K) and small deviations ($\ll1 K$) below
about 20 km. For the test atmosphere with a 100-fold increase in
CO$_2$ (not shown), the stratospheric differences are even larger
(up to 10 K). We interpret these differences in the temperature
profiles as follows:

Firstly, as stated above in section \ref{MRAC}, MRAC does not
differentiate between troposphere and stratosphere, as is the case
for RRTM. That means, spectral bands where H$_2$O or CO$_2$ absorb
only weak are not considered for optical depth calculations in the
stratosphere by RRTM (e.g., bands 6, 12-13 and 15-16). In contrast,
MRAC incorporates the contribution of CO$_2$ and H$_2$O to the
optical depth in these spectral bands. However, this contribution is
usually rather small.

Secondly, and more importantly, the differences between the
stratospheric temperature profiles occur where RRTM has to use a
temperature extrapolation for the absorption cross sections beyond
the limits of its tabulated values.

\begin{figure}[H]
  \includegraphics[width=400pt]{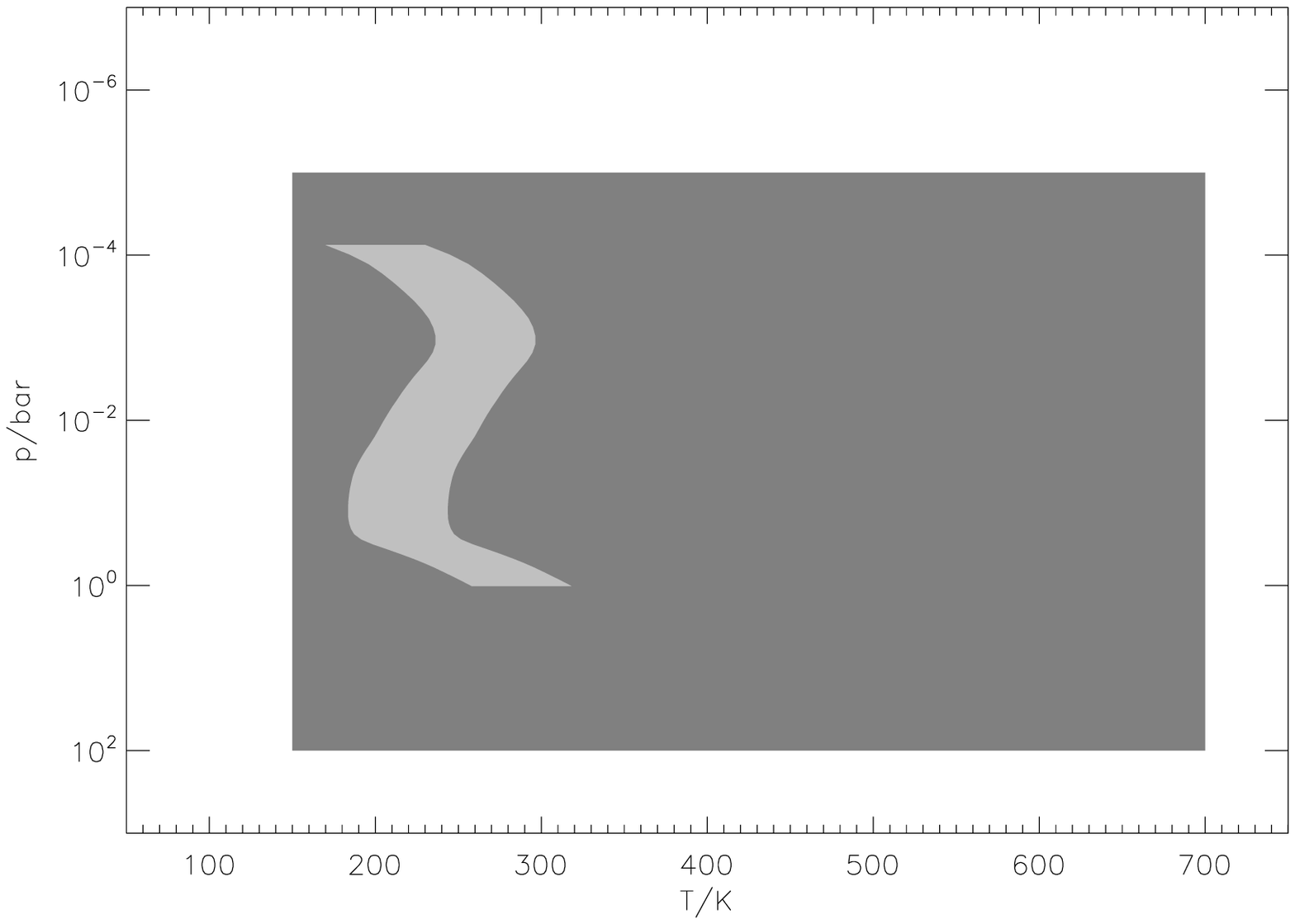}\\
  \caption{}
    \label{range}
\end{figure}

\begin{figure}[H]
  \includegraphics[width=400pt]{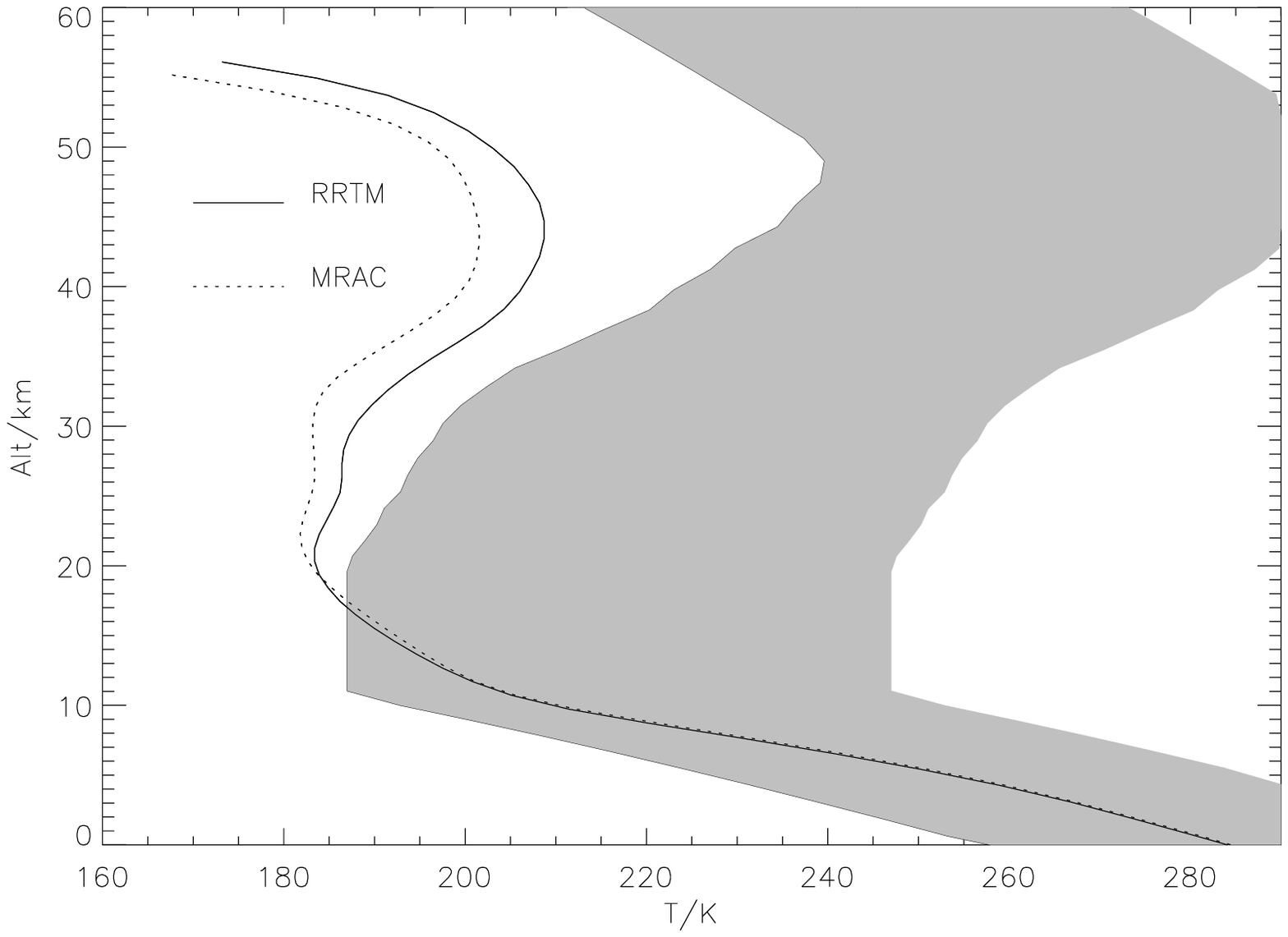}\\
    \caption{}
    \label{limitstable}
\end{figure}

Fig. \ref{limitstable}, shows the temperature profile for our test
atmosphere Earth 1. Also shown is the validity range of RRTM as
already indicated in Fig. \ref{range}. This represents the lower
temperature limit for the tabulated absorption cross sections in
RRTM, as stated above.

As can be seen from Fig. \ref{limitstable}, the calculated
temperature values for the first test atmosphere are below the lower
RRTM validity limit. Therefore, RRTM uses linear extrapolation to
calculate the absorption cross sections. This introduces a large
extrapolation error. On average, the calculated absorption cross
sections are a factor of 2-5 too low, depending on the spectral
band. Sometimes, the extrapolation performed by RRTM even yields
negative absorption cross sections. The interpolation errors in
MRAC, on the contrary, reach only 1-2\% on average.

\begin{figure}[H]
  \includegraphics[width=400pt]{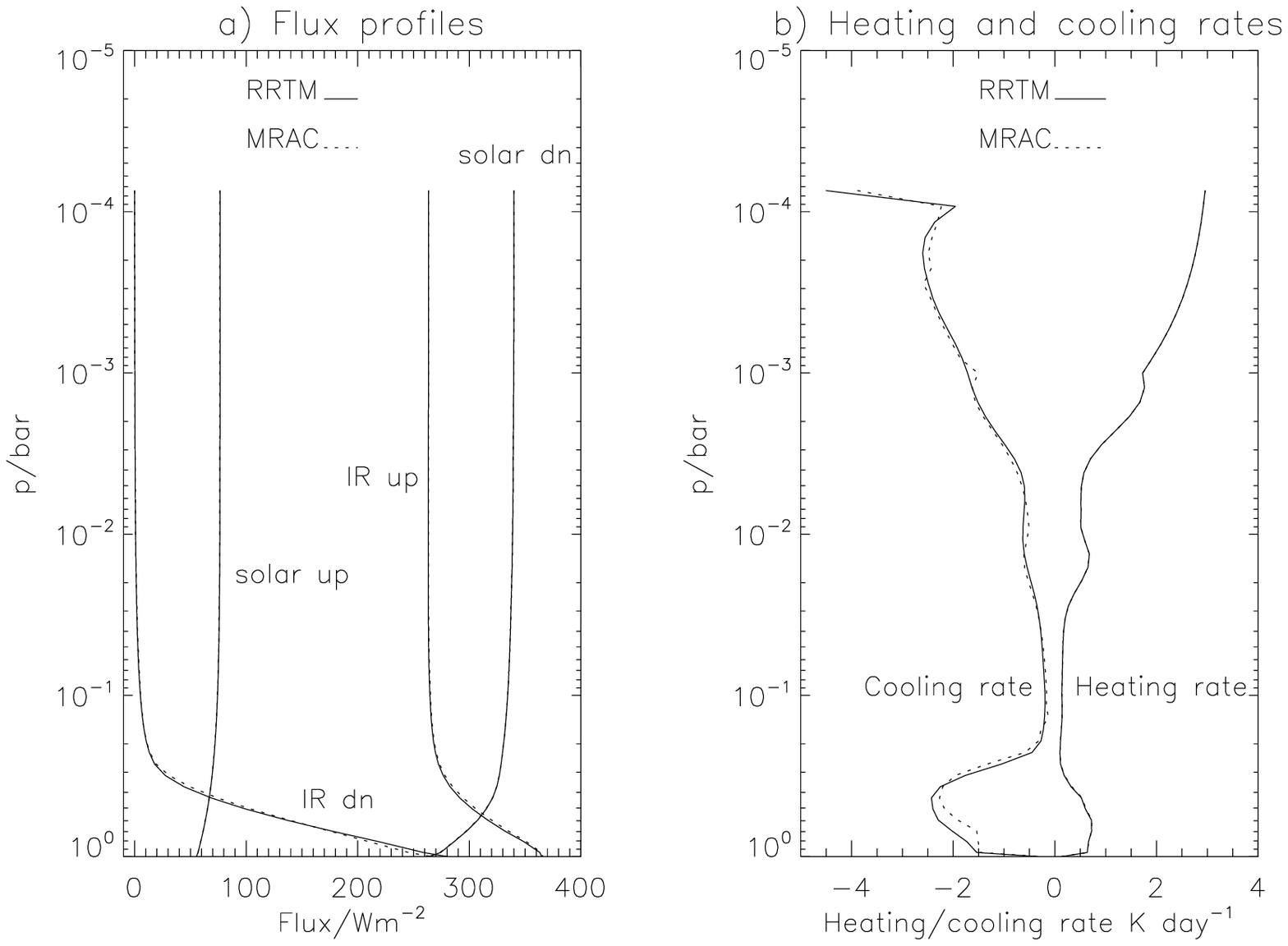}\\
    \caption{}
    \label{fluxdiff}
\end{figure}

Figure \ref{fluxdiff} shows the radiative fluxes and heating and
cooling rates calculated by RRTM and MRAC in the test atmosphere
Earth 1. Solar fluxes and heating rates differ by much less than 1
\%. The thermal down-welling fluxes calculated by RRTM and MRAC show
large differences in the stratosphere below pressures of around
10$^{-2}$-10$^{-3}$ bar, reaching up to a factor of 5 in the upper
stratosphere where pressures are below 10$^{-4}$ bar. However, these
differences are well below 1 W m$^{-2}$, so are not discernible on
the scale in Figure \ref{fluxdiff}. The up-welling fluxes differ by
only about 10 \% in the stratosphere, since they are dominated by
the tropospheric component. Hence, the calculated resulting cooling
rates differ only by small amounts of 0.1-0.4 K day$^{-1}$ and
usually lie within 5-10 \%, especially in the upper stratosphere.

\begin{figure}[H]
  \includegraphics[width=400pt]{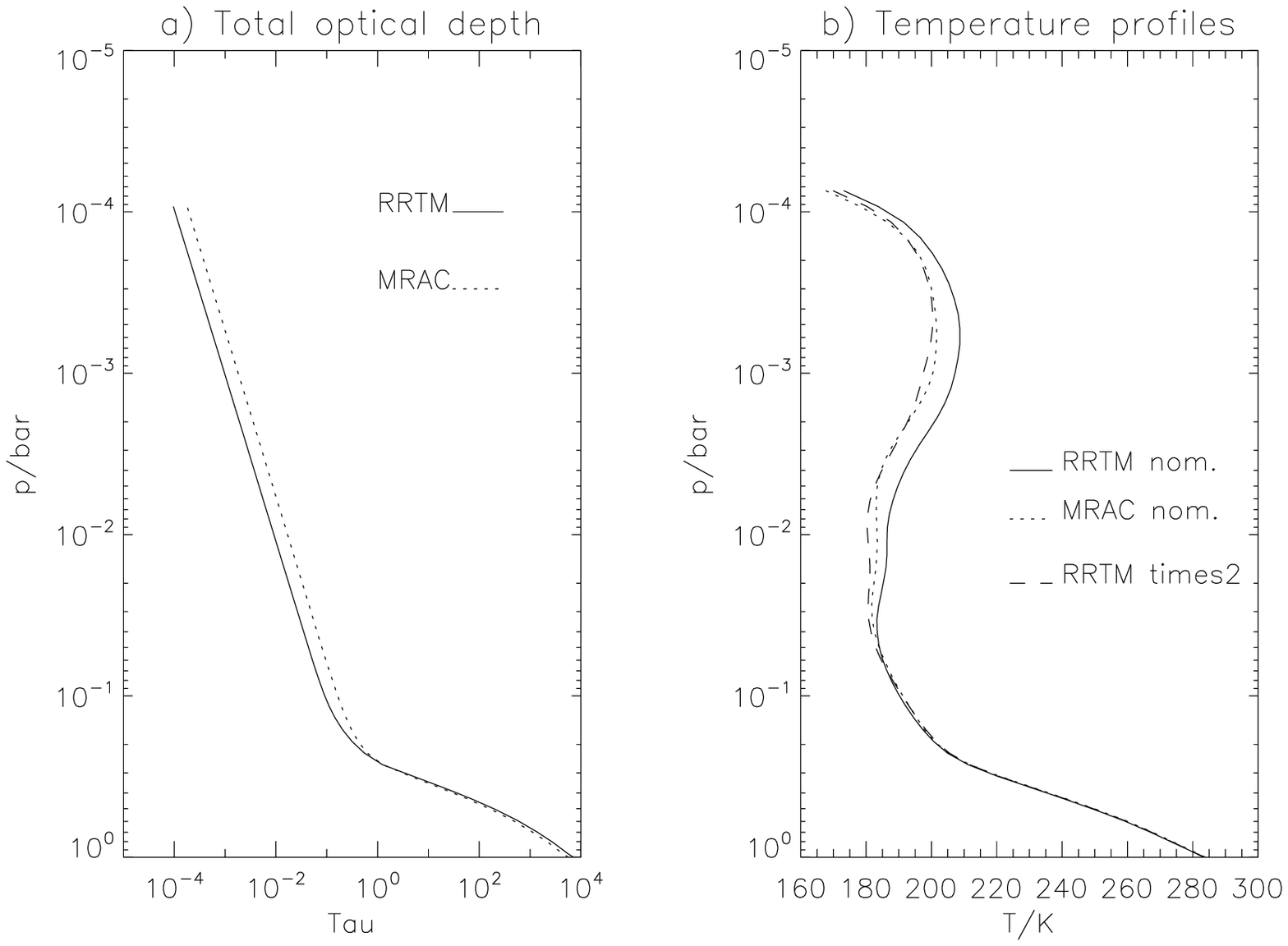}\\
    \caption{}
    \label{vartau}
\end{figure}

In order to assess the sensitivity of the model to errors in the
absorption cross sections (hence, in optical depth and thermal
fluxes), we artificially increased the optical depth in RRTM in the
most important stratospheric band, the CO$_2$ 15$\mu$m fundamental
by factors of 2, 5, 10 and 20, respectively. Fig. \ref{vartau}
quantifies the effect of these sensitivity runs on the temperature
profile. Fig. \ref{vartau} a) shows the total optical depth
calculated by RRTM and MRAC in the validation runs. Clearly, in the
stratosphere, RRTM under-estimates the optical depth, as already
discussed above.


Fig. \ref{vartau} b) shows the temperature profiles for the two
Earth validation runs with MRAC and RRTM, as well as for a run with
RRTM, but increased optical depth by a factor of 2. This factor of 2
is representative of the error in the absorption cross section
calculations in the 15 $\mu$m band due to the required extrapolation
in the $T$-$p$-range, as shown in Figure \ref{range}. It can be seen
that by increasing the optical depth in RRTM artificially, the
temperature profile nearly reproduces the MRAC temperature profile.

These results show that conditions which differ too much from the
Earth's standard atmosphere seem to pose problems for RRTM. This
limitation was already noted in some of the previous studies
performed with RRTM. Clearly, due to the use of an expanded
temperature range, MRAC performs better than RRTM in these
atmospheres.

\section{About the runs}

\label{about}

\subsection{Absorption cross sections}

The absorption cross sections used in the runs performed for this
work (summarized in Tables \ref{runssummary} and \ref{sensruns})
were calculated assuming a N$_2$-CO$_2$-background atmosphere,
consisting of 95\% molecular nitrogen and 5\% carbon dioxide.
According to \citet{kasting1986} and \citet{toth2000}, the foreign
broadening coefficient for water is enhanced by a factor of 2 with
respect to air for CO$_2$ as a broadening gas and by a factor of 1.2
for N$_2$ as a broadening gas. Similarly, for carbon dioxide the
foreign broadening coefficient was enhanced by a factor of 1.3 when
N$_2$ was the broadening gas \citep{kasting1986}. Accounting for the
appropriate mixing ratios then yields an effective enhancement
factor by which the foreign broadening parameter from HiTemp was
multiplied before use in the cross section calculations described
above.

We compared the calculated cross sections of the assumed
N$_2$-CO$_2$-atmosphere (95\% nitrogen and 5\% carbon dioxide) with
cross sections for some major spectral bands corresponding to
different background atmospheres, i.e. different CO$_2$
concentrations. These cross sections were obtained with the
line-by-line radiative transfer model Mirart \citep{schreier2003}.
The cross sections from the 95\%-N$_2$-5\%-CO$_2$-atmosphere agree
within 5 \% for most of the cases studied in this work, although for
the runs with the lowest CO$_2$ concentrations, the agreement
decreased to about 10 \%.
\newline

\subsection{ Model runs}

There were a total number of 12 nominal runs performed as shown in
Table \ref{runssummary}.

We assumed a constant background pressure of 0.77 bar N$_2$ with
variable amounts of CO$_2$. Water vapour contents of the atmosphere
were calculated as described in section \ref{watcalcu}. No other
gases were present in the atmosphere, as in \citet{kasting1987}. For
several values of the solar constant ($S$=0.7, 0.75, 0.8, 0.85, 0.9
and 0.95 present-day value), we increased the CO$_2$ partial
pressure until converged surface temperatures reached 273 K (runs
1-6) and 288 K (runs 7-12), respectively. This procedure is similar
to what was done by \citet{kasting1987}. The solar constant values
were chosen as to loosely correspond to important events throughout
the Earth's history, such as the beginning of the main sequence life
time of the Sun ($S$=0.70, 4.6 Gy ago), the end of the late heavy
bombardment ($S$=0.75, 3.8 Gy ago), the rise of oxygenesis by
cyanobacteria ($S$=0.80, 2.9 Gy ago), the first and the second
oxidation event ($S$=0.85 and $S$=0.90, 2 and 1.3 Gy ago
respectively) and the Cambrian explosion ($S$=0.95, 0.6 Gy ago).
Assigning geological ages to the solar constants used in the model
(see Table \ref{runssummary}) is essential when comparing the
calculated model CO$_2$ concentrations to the available data and
constraints on carbon dioxide. In this work, we chose two
approximations of the solar luminosity $S(t)$ with time. The first
one is from \citet{caldeira1992}, the second approximation is from
\citet{gough1981}. The difference of these two formulations is most
pronounced for earlier time periods before 1-2 Gy ago, although it
is rarely larger than a few percent.

The total surface pressure is always calculated from the relation
$p_{\rm{surf}}=p_{\rm{CO_2}}+p_{\rm{N_2}}+p_{\rm{H_2O}}$. Since we
calculated surface temperatures of 273 K and 288 K, the amount of
surface pressure which arose from the evaporation of water is about
6 mb and 17 mb, respectively. Note that the total surface pressure
in our model runs decreases with time, as we assume less CO$_2$
partial pressure in our model atmospheres. There is no geological
data available on the total surface pressure throughout time,
however, our approach of constant N$_2$ partial pressure is
consistent with assumptions made in previous studies (e.g.,
\citealp{kasting1987}).


\subsection{Parameter variations of surface albedo and relative humidity profile}

Previous works regarding the "faint young Sun problem" have
concentrated on the increased greenhouse effect, as stated above in
the Introduction. We also studied two important parameters affecting
the surface temperature in our model, namely the surface albedo and
the relative humidity (RH) distribution.

The importance of cloud coverage for the surface temperature, e.g.
on early Mars, has been studied by \citet{mischna2000}. The effect
in their model was quite large, yielding surface temperatures from
220 K to 290 K, depending on cloud optical depth and cloud location.
In our model, the surface albedo simulates the presence of clouds,
as stated above in the description of our model.

Water vapour is a very effective greenhouse gas. In contrast to
CO$_2$, its content in the troposphere (where 99 \% of the column
resides) is controlled by the hydrological cycle (ocean reservoir,
evaporation, subsequent condensation and precipitation) which is
very sensitive to temperature. A critical parameter in climate
models is the relative humidity parametrization, which clearly has a
potentially large impact on the water content in the atmosphere and
hence on the resulting greenhouse effect.

In addition to the model runs described above, we therefore
performed runs varying surface albedo and relative humidity
profiles. These runs are based on runs 3 and 4 of Table
\ref{runssummary}, i.e. solar constants of $S$=0.8 and $S$=0.85.
Table \ref{sensruns} summarizes the parameter variations.


In the runs from Table \ref{runssummary}, the surface albedo is
normally set to $A$=0.21 (see Table \ref{paraminput}). Now, we
assume surface albedos of $A$ $\pm$10 \%, i.e. 0.19 and 0.23,
respectively.

\begin{figure}[H]
  \includegraphics[width=400pt]{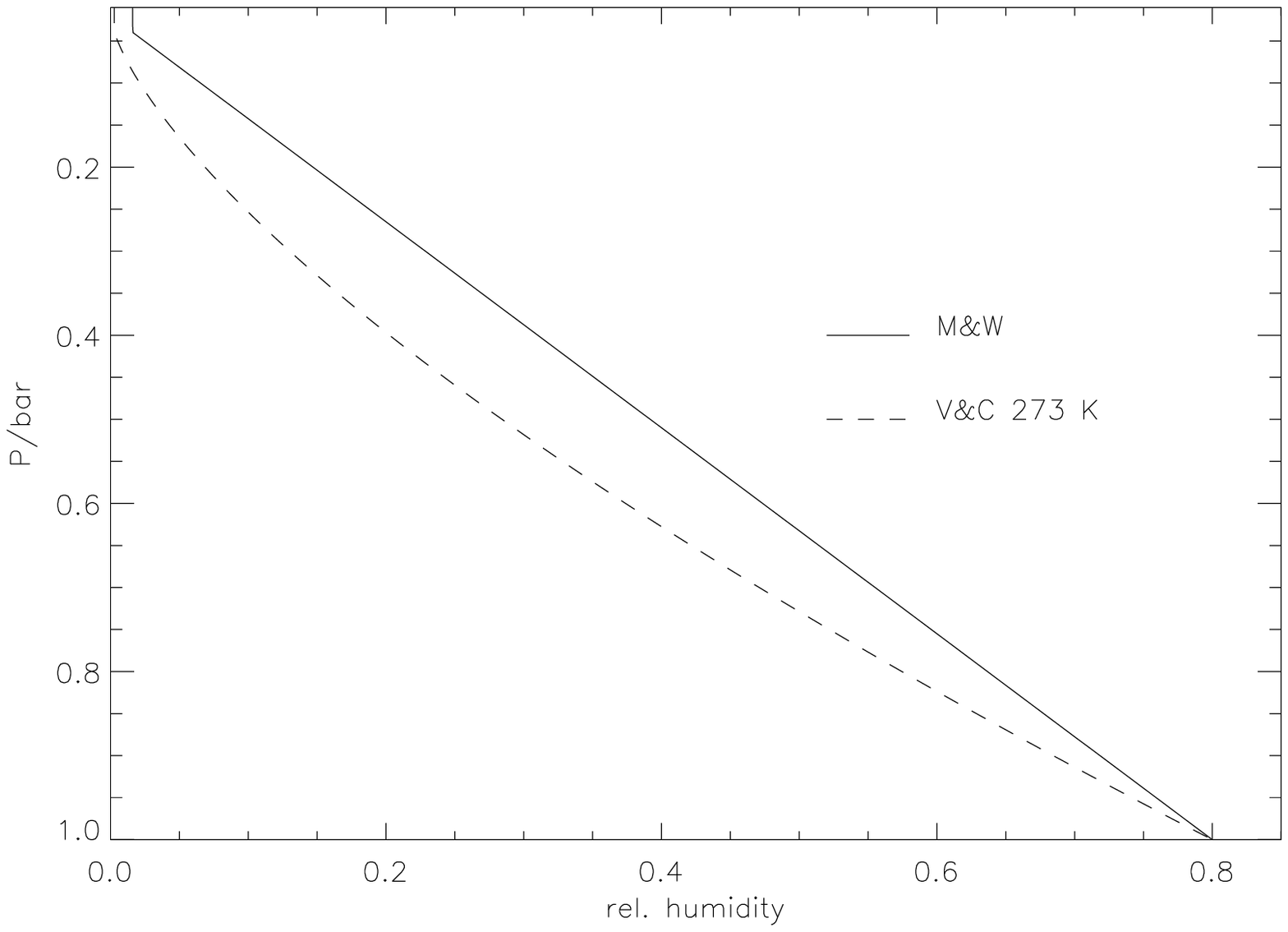}\\
      \caption{}
    \label{relhumdif}
\end{figure}

The relative humidity profile used for the calculation of water
vapour concentrations in Table \ref{runssummary} follows the
approach of \citet{manabewetherald1967} (referred to as RH=MW, see
section \ref{watcalcu}). Here, for the additional parameter studies,
we used three different RH profiles. The first one assumed a
saturated troposphere (RH=1). The second profile used RH=0, i.e.
water was removed from the atmosphere. These two profiles represent
the lower and upper limits of the atmospheric water content. The
third RH profile is a more realistic RH profile. It is based on a
temperature correction to the RH profile of
\citet{manabewetherald1967} which was first proposed by
\citet{cess1976} and later used by \citet{vardavas1985}.:

\begin{equation}\label{relhumcess}
R=R_{\rm{surface}}\cdot R_{\rm{mw}}^{1-0.03(T_S-288)}
\end{equation}

where $R_{\rm{surface}}\cdot R_{\rm{mw}}$ is the relative humidity
distribution from eq. \ref{manabewetherald}. This profile is
referred to as RH=C. Figure \ref{relhumdif} shows the difference
between the two RH profiles from \citet{manabewetherald1967} and
\citet{vardavas1985}. The total amount of water vapour is reduced by
about 20 \% by using the temperature correction of
\citet{vardavas1985} compared to \citet{manabewetherald1967}.


\section{Results and Discussion}

\label{showresults}

\subsection{Examples of thermal structure and water profiles}

\begin{figure}[H]
  \includegraphics[width=400pt]{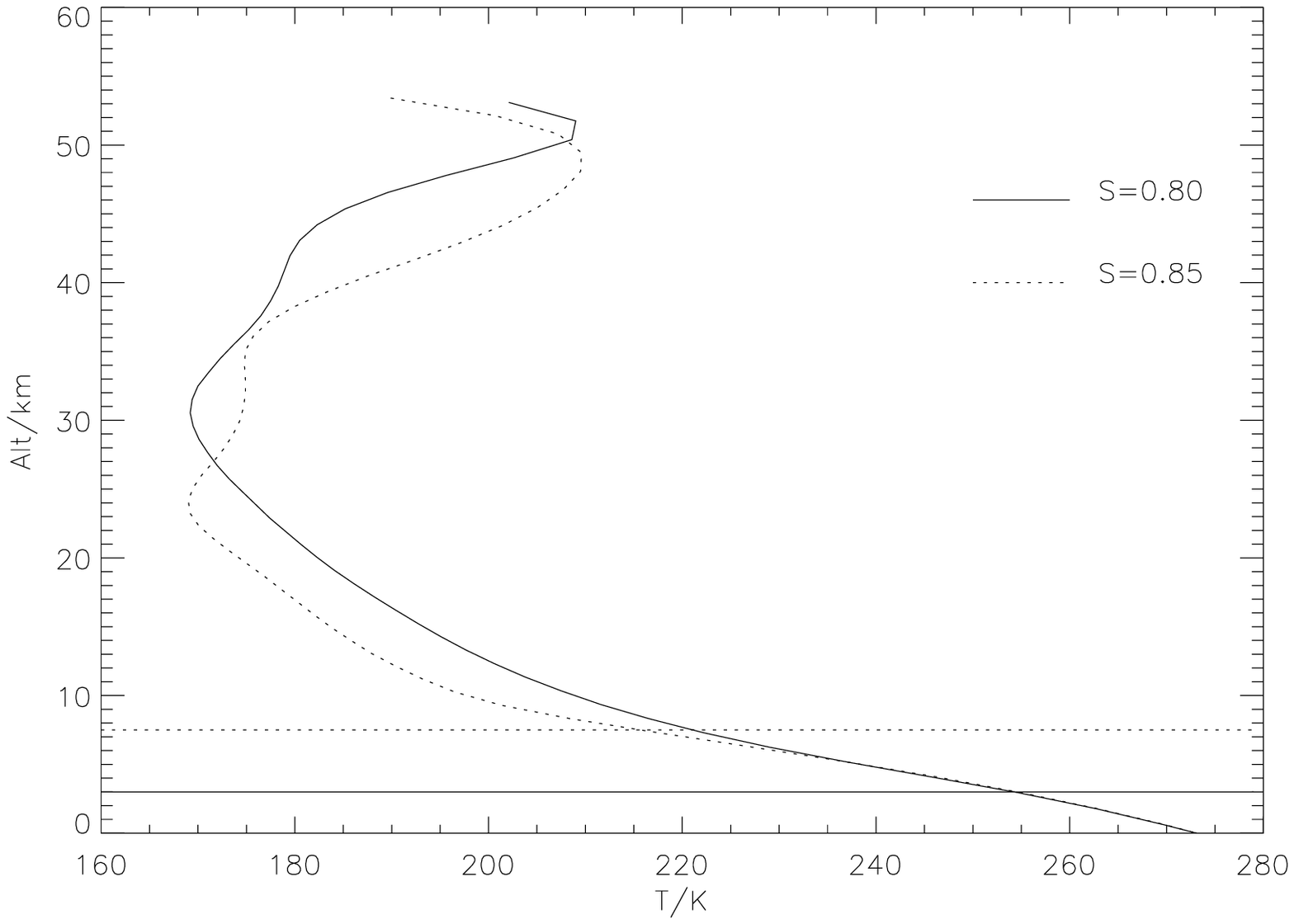}\\
      \caption{}
    \label{tprofile}
\end{figure}

Figure \ref{tprofile} shows the temperature profiles for run 3
(plain line) and run 4 (dotted line). Run 3 ($S=0.80$) considered 20
mb CO$_2$ partial pressure, run 4 ($S=0.85$) roughly 3 mb (see Table
\ref{runssummary}). These two runs were chosen because they
represent interesting points in time, such as the advent of
cyanobacteria (run 3) and the first oxidation event (run 4).


Clearly, in the troposphere the temperature differences for the two
runs are rather small. In the lower to middle stratosphere (from
around 10 to 25 km),  run 4 (dashed line) shows lower temperatures
than run 3 (plain line). In this region, absorption of solar
radiation by CO$_2$ is the dominant heating process, eventually
causing a temperature inversion from the middle stratosphere upwards
(seen for both runs in Fig. \ref{tprofile} for higher altitudes
above 25-30 km). In the upper stratosphere (above 25-30 km), where
CO$_2$ radiative cooling via the 15 $\mu$m fundamental band sets in,
run 3 shows lower temperatures than run 4. This reflects the lower
flux and higher CO$_2$ concentrations in run 3 compared to run 4.

However, as already noted by \citet{Seg2003}, the actual strength of
the temperature inversion is not well determined due to less
accurate transmittance calculations in the solar code and may well
turn out to be a mere numerical problem rather than reflecting
physical processes.

Also indicated in Fig. \ref{tprofile} are the convective zones for
both runs. The convective zone for the $S=0.8$ case only reaches up
to about 3 km altitude, whereas in the $S=0.85$ case, it extends
already to about 7.5 km, which is closer to the present-day,
latitude-dependent value of 10-20 km. However, the lapse rate
remains close to the adiabatic value for several kilometres after
entering the regime of radiative equilibrium (above the dot-dashed
lines in Fig. \ref{tprofile}).

\begin{figure}[H]
  \includegraphics[width=400pt]{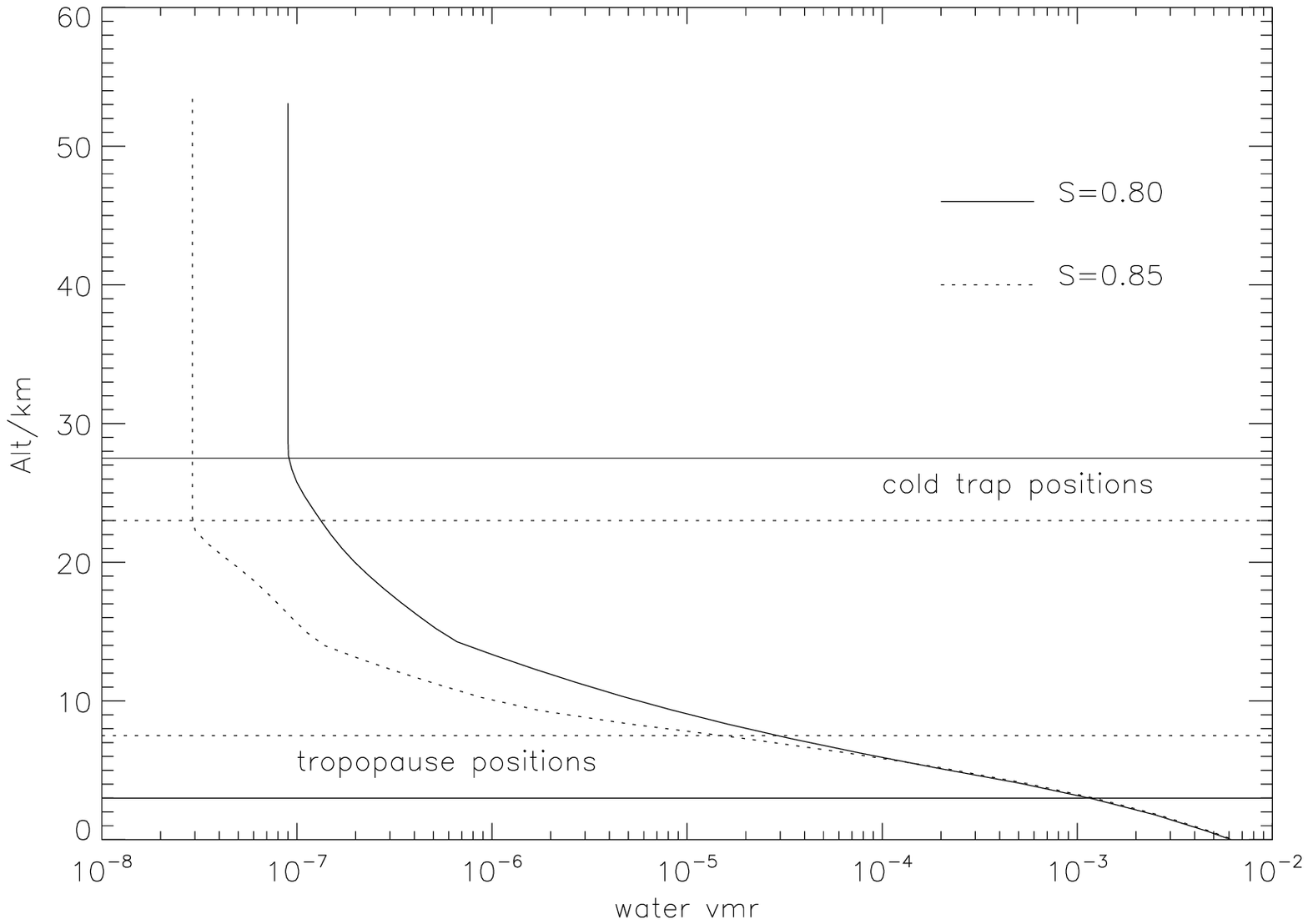}\\
      \caption{}
    \label{wprofile}
\end{figure}

Figure \ref{wprofile} is similar to Fig. \ref{tprofile}, but it
shows the resulting water profiles. There is considerably less water
in the stratosphere for the $S=0.85$ case due to enhanced
condensation. The cold trap position is indicated, as well as the
tropopause position, i.e. the boundary between the convective and
non-convective regime.


Figures \ref{tprofile} and \ref{wprofile} illustrate some
interesting features regarding the thermal structure of the
atmosphere. The cold trap, i.e. the highest point up to which water
is allowed to condense out in the model, is no longer located near
the tropopause, as is the case in the atmosphere of modern Earth. It
is still associated with a temperature inversion, but as this
inversion occurs at higher altitudes, cold trap and tropopause
(i.e., the boundary between convective and non-convective layers)
are located at noticeably different altitudes. It seems that the
observation that tropopause, cold trap and temperature inversion
occur at approximately the same height in the modern Earth's
atmosphere is somewhat coincidental.

\subsection{Climatic constraints on CO$_2$ partial pressures}

\begin{figure}[H]
    \includegraphics[width=400pt]{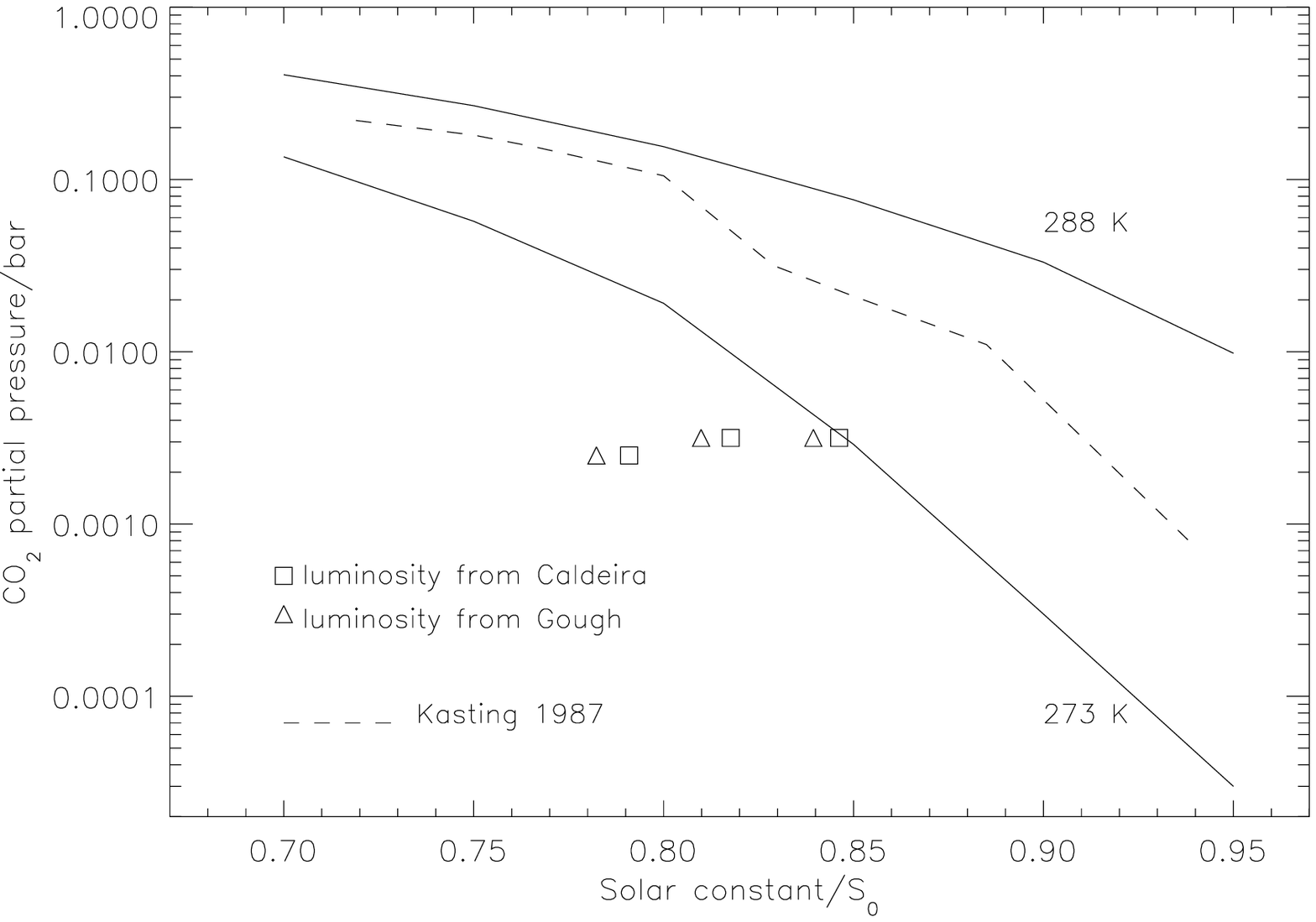}\\
    \caption{}
    \label{minin2data}
\end{figure}

The results of the 12  nominal  runs (see Table 4) are summarized in
Fig. \ref{minin2data}. The lower plain line indicates the CO$_2$
partial pressures corresponding to calculated surface temperatures
of 273 K (runs 1-6), the upper plain line shows the partial
pressures for surface temperatures of 288 K (runs 7-12). For
comparison, the dashed line shows the partial pressures as
calculated by \citet{kasting1987}. Fig. \ref{minin2data} shows that
our new radiative scheme requires considerably less CO$_2$ (about a
factor of 2-15) to achieve an ice-free surface than the study of
\citet{kasting1987}.


Note that the radiative transfer scheme used by \citet{kasting1987}
was applicable to this type of atmospheres, as is the one we used.

Also indicated in Fig. \ref{minin2data} are the upper limits of
CO$_2$ partial pressures as inferred from sedimentary data.

Several attempts have been made to constrain the CO$_2$ content of
the early Earth atmosphere. For example, among others,
\citet{rye1995}, \citet{hessler2004} and \citet{towe1985} have
placed upper limits on CO$_2$ partial pressures during different
periods of the mid-Archaean to early Proterozoic of approximately
10-100 PAL (Present Atmospheric Level), depending on temperature.
These limits were based on experimental sediment data and on
biological considerations. \citet{rye1995} and \citet{hessler2004}
argued for upper limits on CO$_2$ partial pressures based on the
observations that in ancient Archaean rocks, siderites are missing.
\citet{towe1985} stated atmospheric upper levels of CO$_2$ based on
the observation that anaerobic photosynthesis and nitrogen fixation
(which are both believed to be evolutionary ancient) are
incompatible with high CO$_2$ concentrations.

Previous climate studies of the early Earth's atmosphere calculated
very high CO$_2$ partial pressures in order to raise the surface
temperature above 273 K. The model CO$_2$ values were generally
above 50 mb well into Proterozoic age (e.g., Kasting 1987, see also
Fig. \ref{minin2data}). However, the experimental data from
sediments and biology (see above) were of the order of several mb,
i.e. an order of magnitude lower. This contradiction is known as the
"faint young Sun problem".

Fig. \ref{minin2data} however shows that for the late Archaean
(solar constant S=0.85), our new results are compatible with the
paleosol records, hence the "faint young Sun problem" might be
resolved for this time period.

One should note that the absence of methane, ozone, ammonia or other
greenhouse gases in our results implies that we calculated only a
lower limit for surface temperatures. Photochemical models of the
anoxic Archaean and low-oxygen Proterozoic atmospheres have shown
that methane could build up to concentrations of the order of
10$^{-3}$ \citep{pavlov2003} which could also have contributed to
the greenhouse effect. \citet{haqq2008} proposed higher hydrocarbons
such as C$_2$H$_6$ in a methane-rich atmosphere as radiative gases.
Those effects were not investigated here. Our studies imply,
however, that much less to none additional greenhouse gases are
required to warm the early Earth.

\subsubsection{Effect of parameter variations of surface albedo and relative humidity}

Table \ref{sensruns} summarizes the results of the parameter
variations. Shown are the values of CO$_2$ partial pressure required
to obtain the desired surface temperature of 273 K.

Upon lowering the surface albedo value from 0.23 to 0.19, the
partial pressure of CO$_2$ required to keep the surface at 273 K is
reduced from 28.5 mb to 11.0 mb (at $S$=0.80) and from 5.5 mb to 1.5
mb (at $S$=0.85). This is a lowering of the amount of CO$_2$ by
factors of 2.6 and 3.7, respectively.

When changing the RH profile from saturated (RH=1) to water-free
conditions (RH=0), the necessary partial pressure of CO$_2$ has to
be increased from 4.9 mb to 266.8 mb ($S$=0.80) and from 0.6 mb to
180.6 mb ($S$=0.85). However, like expected, the comparison between
the more realistic RH profiles of \citet{manabewetherald1967} and
\citet{cess1976} yields much smaller differences. When changing the
RH profile from RH=MW to RH=C, CO$_2$ partial pressures must be
increased from 19.1 mb to 27.3 mb ($S$=0.80) and from 2.9 mb to 4.9
mb ($S$=0.85). These differences amount to 43 \% and 69 \%,
respectively, which is much lower than the differences obtained by
varying the surface albedo.


These parameter variations demonstrated that the surface albedo,
hence clouds, is a very important parameter in view of the surface
temperature. The results obtained imply that the latter is probably
more important than the RH profile, although the RH profile has to
be incorporated more consistently to accurately constrain CO$_2$
partial pressures for the "faint young Sun" problem. This calls for
a more elaborated 1D model incorporating clouds and cloud formation
as was done by, e.g., \citet{Cola2003} for early Mars. In the
future, we plan to add clouds to our code.

The results of the parameter variations, however, did not
significantly change our main conclusion, namely that the CO$_2$
values for the late Archaean calculated by our improved model are
consistent with observational data. The CO$_2$ values obtained for
the parameter variations range between 1.5-5.5 mb of partial
pressure. This is still close to the values inferred from the
paleosol record.

\section{Summary}

\label{conclusions}

In this work, we addressed the "faint young Sun problem" of the
ice-free early Earth. In order to do this, we applied a
one-dimensional radiative-convective model to the atmosphere of the
early Earth. Our model included updated absorption coefficients in
the thermal radiative transfer scheme.

The validations done for the new radiative transfer scheme have been
described. The new scheme is found to perform significantly better
than
 a previous scheme under conditions deviating from the modern
Earth atmosphere.

We then studied the effect of enhanced carbon dioxide concentrations
and parameter variations of the surface albedo and the relative
humidity profiles on the surface temperature of the early Earth with
the improved model.

Our new model simulations suggest that the amount of CO$_2$ needed
to keep the surface of the early Earth from freezing is
significantly less (up to an order of magnitude) than previously
thought (see Figure \ref{minin2data}).

For the late Archaean and early Proterozoic period (around 2-2.5 Gy
ago), the calculated amount of CO$_2$ (2.9 mb partial pressure)
which is needed to obtain a surface temperature of 273 K is
compatible with the amount inferred from geological data, contrary
to previous studies (see Figure \ref{minin2data}). The apparent
contradiction between model constraints on CO$_2$ and sediment data
disappears for this time period.

Upon varying model parameters such as the surface albedo and
relative humidity profile, we found this conclusion to be robust.
The calculated CO$_2$ partial pressures for the late Archaean
(1.5-5.5 mb) are still consistent with the geological evidence.

\section*{Acknowledgements}

We are grateful to Jim Kasting and Eli Mlawer for useful discussion
while creating the new radiation scheme. Furthermore, we are
grateful to Viola Vogler for help in doing some of the plots in this
paper.

We thank the two anonymous referees for their constructive remarks
which helped to improve and clarify this paper.

\newpage

\bibliographystyle{elsart-harv}
\bibliography{v2}

\newpage
\linespread{1.}

\section*{Figure captions}

Figure \ref{range}:

Range of $T$-$p$ values used to obtain absorption cross sections in
the two radiative schemes RRTM (light grey) and MRAC (dark grey).
\newline

Figure \ref{kdislacis}:

Cumulative $k$ distribution for parts of the 6.3 $\mu$m H$_2$O
fundamental band, T=240 K and p=10 mb: Comparison between
\citet{lacis1991} (plain line) and
  our algorithm (dotted).
\newline

Figure \ref{kdismlawer}:

Cumulative $k$ distribution for parts of the 15 $\mu$m CO$_2$
fundamental band between
  630 and 700 cm$^{-1}$, T=260 K and p=507 mb: Comparison between \citet{Mlawer1997} (plain line)
  and our algorithm (dotted).
\newline

Figure \ref{MRACvalid}:

Comparison of validation temperature profiles (RRTM plain line, MRAC
dotted line).
\newline

Figure \ref{limitstable}:

Limits of the RRTM temperature grid. Atmospheric conditions as for
the standard CO$_2$ case in fig. \ref{MRACvalid}, RRTM profile
(plain), MRAC (dotted). The shaded area designates the validity
range of RRTM as already indicated in fig. \ref{range}.
\newline

Figure \ref{fluxdiff}:

Flux profiles (thermal and solar up- and down-welling fluxes) (left
panel) and heating and cooling rates (right panel) calculated by the
two radiative schemes in the test atmosphere.
\newline

Figure \ref{vartau}:

Effect of variations in the optical depth on the temperature profile
calculated by RRTM: Left panel, total optical depth, right panel,
temperature profiles.
\newline

Figure \ref{relhumdif}:

Differences of the two relative humidity profiles used for the
parameter variations
\newline

Figure \ref{tprofile}:

Temperature profiles for runs with solar constants of $S=0.8$ (run
3, plain line) and $S=0.85$ (run 4, dotted); the background pressure
is 0.77 bar of N$_2$. The convective zones are indicated for both
runs.
\newline

Figure \ref{wprofile}:

Water profiles for runs with solar constants of $S=0.8$ (run 3,
plain line) and $S=0.85$ (run 4, dashed); the background pressure is
0.77 bar of N$_2$. Cold trap positions and tropopause positions are
indicated for both runs.
\newline

Figure \ref{minin2data}:

Minimal values of CO$_2$ partial pressure required to obtain chosen
surface temperatures of 273 K and 288 K at a fixed N$_2$ partial
pressure for different solar constants: shown are the curves for the
new model (plain lines, upper line 288 K, lower line 273 K) and the
model of \citet{kasting1987} (dashed)). Symbols are included which
represent the upper CO$_2$ limits derived from the sediment record
($\Box$: age conversion by \citet{caldeira1992}, $\triangle$: age
conversion by \citet{gough1981}.

\newpage
\section*{Tables}

\input{table1}
\newpage

\input{table2}
\newpage

\input{table3}
\newpage

\input{table4}
\newpage

\section*{Figures}

\newpage

\newpage

\newpage

\newpage

\newpage

\newpage

\newpage

\newpage

\newpage

\newpage

\end{document}

%% file: table1.tex
\begin {table}[H]

\center{

\caption{Contribution of model species to the temperature profile
via radiative transfer for solar or thermal radiation, adiabatic
lapse rate formulations or heat capacity contributions (x: active
species, -: inactive species)}

\begin {tabular}{|l|c|c|c|c|c|c|}

\hline

Gas& Solar &Rayleigh & Thermal &  Continuum & Lapse Rate & Heat Cap.\\
\hline

N$_2$& - &x &  - &  - & -& x\\
\hline

H$_2$O& x &- & x  &  x & x& -\\
\hline

O$_2$& - &x &  - &  - & -& x\\
\hline

Ar   & - & - &  - &  - & -& x\\
\hline

CO$_2$& x &x &  x &  x & -& x\\
\hline

CO   & - & - &  x &  - & -& x\\
\hline

\end{tabular}

 \label{thermalinput}

}

\end{table}

%% file: table2.tex
\begin {table}[H]

\caption{Initial values, boundary conditions and parameters in the
model (IV: initial value, BC: boundary condition, P: parameter)}

\begin{centering}

\begin{tabular}{|l|l|c|}

\hline

Quantity& Value& Type\\

\hline

$T_0$-profile& US Standard 1976& IV\\

\hline

TOA incident flux& solar spectrum& BC\\

\hline

TOA $p_0$& 6.6$\cdot 10^{-5}$ bar& P\\

\hline


Surface albedo& 0.21& P\\

\hline

Zenith angle& 60$^\circ$& P\\

\hline

\end{tabular}

 \label{paraminput}

\end{centering}

\end{table}

%% file: table3.tex
\begin {table}[H]

\center{

\caption{Summary of model runs performed for this work (partial
pressures $p_{\rm{N_2}}$ and $p_{\rm{CO_2}}$ and surface pressure
$p_{\rm{surf}}$ are in mb, surface temperature $T_{\rm{surf}}$ in
K)\newline}

\begin {tabular}{|c|c|c|c|c|c|}

\hline

Runs & solar constant $S$& $p_{\rm{surf}}$ & $p_{\rm{CO_2}}$ & $p_{\rm{N_2}}$ & $T_{\rm{surf}}$\\

\hline

 1 & 0.70 & 911& 135.3& 770 & 273\\

\hline
 2 & 0.75 & 833& 57.2 & 770 & 273\\

\hline

 3 & 0.80 &795 & 19.1  & 770 & 273\\

\hline

 4 & 0.85 & 779 & 2.9 & 770 & 273\\

\hline

 5 & 0.90 & 776& 0.3 & 770 & 273\\

\hline
 6 & 0.95 & 776&0.03 & 770 & 273\\

\hline

\hline
 7 & 0.70 &1192 &405.6 & 770 & 288\\

\hline
 8 & 0.75 &1055 & 268.1 & 770 & 288\\

\hline

 9 & 0.80 & 942 & 155.0  & 770 & 288\\

\hline

 10 & 0.85 & 863 & 76.2 & 770 & 288\\

\hline

 11 & 0.90 & 820 & 33.1 & 770 & 288\\

\hline
 12 & 0.95 & 797& 9.8 & 770 & 288\\

\hline
\end{tabular}

 \label{runssummary}

}

\end{table}

%% file: table4.tex
\begin {table}[H]
 \center{

\caption{Summary of sensitivity runs performed (RH profiles: MW
\citet{manabewetherald1967}, C \citet{cess1976}). Partial CO$_2$
pressure $p_{\rm{CO_2}}$ to reach 273 K in mb.\newline}

\begin {tabular}{|c|c|c|c|c|c|}

\hline

Run number & solar constant $S$& RH profile & Surface albedo & $T_{\rm{surf}}$& $p_{\rm{CO_2}}$ \\

\hline

 1a & 0.80 & MW& 0.19&  273& 11.0\\

 \hline

 2a& 0.80 & MW & 0.21 &  273 & 19.1\\

\hline
 3a & 0.80 & MW& 0.23&  273 & 28.5\\

\hline

 4a& 0.85 & MW& 0.19&  273 & 1.5\\

\hline

 5a& 0.85 & MW & 0.21 &  273 & 2.9\\

\hline
 6a & 0.85 & MW& 0.23&  273 & 5.5\\

\hline \hline

 7a& 0.80 & MW & 0.21 &  273 & 19.1 \\

\hline
 8a & 0.80 & C& 0.21&  273 & 27.3\\

\hline

 9a & 0.80 & 1& 0.21&  273 &4.9 \\

\hline

 10a & 0.80 & 0& 0.21&  273& 266.8\\

\hline

 11a& 0.85 & MW & 0.21 &  273& 2.9\\

\hline
 12a & 0.85 & C& 0.21&  273& 4.9\\

\hline

 13a & 0.85 & 1& 0.21&  273 &0.6\\

\hline

 14a & 0.85 & 0& 0.21&  273 &180.6\\

\hline

\end{tabular}

 \label{sensruns}

}
\end{table}